\documentclass[nofootinbib,prd,twocolumn,superscriptaddress,showpacs]{revtex4}

\usepackage{empheq}
\usepackage{amsfonts}
\usepackage{soul}
\usepackage[normalem]{ulem}
\usepackage[english]{babel}
\usepackage[latin1]{inputenc}
\usepackage{lastpage} 
\usepackage{amssymb}
\usepackage{bm}
\usepackage{graphicx}
\usepackage[hang,normal]{subfigure}
\usepackage{enumerate}
\usepackage[usenames,dvipsnames]{color}
\newcommand{\eqeqref}[1]{Eq.~\eqref{#1}}

\newcommand{\secref}[1]{Section~\ref{#1}}

\def\be{\begin{equation}}
\def\ee{\end{equation}}
\def\bea{\begin{eqnarray}}
\def\eea{\end{eqnarray}}

\begin{document}
\title{Geometrothermodynamic model for the evolution of the Universe}
\date{\today}

\author{Christine Gruber}
\email{christine.gruber@correo.nucleares.unam.mx}
\affiliation{Instituto de Ciencias Nucleares, Universidad Nacional Aut\'onoma de M\'exico,\\ AP 70543, M\'exico, 
DF 04510, Mexico}

\author{Hernando Quevedo}
\email{quevedo@nucleares.unam.mx}
\affiliation{Instituto de Ciencias Nucleares, Universidad Nacional Aut\'onoma de M\'exico,\\ AP 70543, M\'exico, 
DF 04510, Mexico}
\affiliation{Dipartimento di Fisica and ICRA, Universit\`a di Roma ``La Sapienza'', I-00185 Roma, Italy}

\begin{abstract}
Using the formalism of geometrothermodynamics to derive a fundamental thermodynamic equation, we construct a 
cosmological model in the framework of relativistic cosmology. In a first step, we describe a system without 
thermodynamic interaction, and show it to be equivalent to the standard $\Lambda$CDM paradigm. The second step 
includes thermodynamic interaction and produces a model consistent with the main features of inflation. With 
the proposed fundamental equation we are thus able to describe all the known epochs in the evolution of our 
Universe, starting from the inflationary phase. 
\end{abstract}
\pacs{98.80.Cq; 98.80.-k; 95.30.Tg; 02.40.-k}
% Explainations: Inflationary universe; Cosmology; Thermodynamics in Astrophysics; Differential geometry

\maketitle

%%%%%%%%%%%%%%%%%%%%%%%%%%%%%%%%%%%%%%%%%%%%%%%%%%%%%%%%%%%%
%%%%%%%%%%%%%%%%%%%%%%%%%%%%%%%%%%%%%%%%%%%%%%%%%%%%%%%%%%%%%
\section{Introduction}

The standard cosmological model of general relativity states that the evolution of our Universe is described 
by the $\Lambda$CDM model, and postulates that there must have been an inflation era early in the history in which the 
vacuum energy dominated over other forms of energy density such as baryonic matter, dark matter or radiation 
\cite{lcdm}.  All the physical phases contained in the $\Lambda$CDM scenario and inflation are characterized 
by particular equations of state which usually relate the corresponding energy density with the pressure. There 
are many ways to impose or derive a particular equation of state. Recently in \cite{abcq12}, we proposed a method 
in which the equations of state are derived from a fundamental thermodynamic equation which, in turn, can be obtained 
by using the formalism of geometrothermodynamics (GTD). One of the central ideas of GTD \cite{quev07} is that any 
thermodynamic system is represented by its corresponding space of equilibrium states ${\cal E}$. Moreover, the 
thermodynamic properties of the system can be represented in terms of the geometric properties of the equilibrium 
space -- e.g., the curvature of the equilibrium space can be interpreted as 
a measure of the thermodynamic interaction and curvature singularities represent phase transitions. This has been 
shown to be true for a large number of thermodynamic systems (see \cite{qqs16b} and references therein). 

If the system contains $n$ different thermodynamic degrees of freedom (number of independent thermodynamic 
variables that are necessary for its description), the corresponding equilibrium space is $n-$dimensional whose 
geometric properties are completely determined by an $n-$dimensional metric $g$. In thermodynamic geometry, the 
components of the metric $g$ are usually identified with the second derivatives of a particular thermodynamic 
potential which can be chosen as the internal energy (Weinhold metric), the entropy (Ruppeiner metric), or in 
terms of a partition function (Fisher-Rao metric). These particular Hessian metrics have been extensively applied 
to explore the properties of a large number of thermodynamic systems and also in connection with information theory 
\cite{amari85,weibook,rup14}. The approach of GTD is different. The metric $g$ is chosen such that it is invariant 
with respect to Legendre transformations. This is an additional condition which is necessary in GTD in order to 
incorporate the well-known property of the Legendre invariance of classical thermodynamics into the geometric 
description, i.e., the independence of the physical behaviour of the system on the choice of thermodynamic potential. 
Furthermore, in order to consider a Legendre transformation as a coordinate transformation, in GTD it is necessary 
to introduce the $(2n+1)-$dimensional phase space ${\cal T}$ which is endowed with a contact one-form $\Theta$ and 
a metric $G$. A Legendre transformation in ${\cal T}$ can be expressed as a coordinate transformation 
$Z^A \rightarrow \tilde Z ^A$ ($A=0,1,...,2n)$ where the Jacobian is the unity matrix \cite{quev07}. The 
contact one-form can be written in a canonical manner as $\Theta = d\Phi - \delta_{ab}I^a dE^b$ ($a=1,...,n,)$ with 
$\delta_{ab}={\rm diag}(1,1,...)$, where the coordinates have been chosen as  $Z^A=\{\Phi, E^a, I^a\}$. Whereas the 
contact form is Legendre invariant in 
the sense that under a Legendre transformation $Z^A \rightarrow \tilde Z ^A$ it behaves as $\Theta \rightarrow 
\tilde\Theta = d\tilde \Phi -\delta_{ab} \tilde I^a d\tilde E^a$, the components of the metric $G$ are in general 
not Legendre invariant. Nevertheless, it is possible to find particular solutions which preserve Legendre 
invariance. For instance, the metric (summation over all repeated indices implied) 
\be
  \label{GIIIL}
  G  =(d\Phi - I_a d E^a)^2  + \Lambda \left(E_a I_a \right)^{2k+1}  d E^a   d I^a \,,
\ee
where $ I_a = \delta_{ab} I^b$, $k$ is an integer, and $\Lambda=\Lambda(Z^A)$ is an arbitrary non-zero Legendre-invariant 
function, is the most general metric found so far which is invariant with respect to partial and total 
Legendre transformations. 

The connection between the phase space ${\cal T}$ and the equilibrium space ${\cal E}$ is determined through 
an embedding map $\varphi: {\cal E} \rightarrow {\cal T}$ such that its pullback $\varphi^*$ satisfies the condition 
$\varphi^*(\Theta) = 0$. In particular, we can choose $\{E^a\}$ as the coordinates of ${\cal E}$, in which case the 
pullback condition reads 
\be
  d\Phi = I_a d E^a \,.
\ee
This is the general form of the first law of thermodynamics where $\Phi$ is the thermodynamic potential, the $E^a$ 
are the extensive variables and the $I_a$ the intensive variables dual to $E^a$, i.e., $I_a = \Phi_{,a}\equiv 
\frac{\partial \Phi}{\partial E^a}$.  Furthermore, the pullback induces a canonical metric $g$ on the equilibrium 
space ${\cal E}$, 
\be
  g = \Lambda\, (E_a\Phi_{,a})^{2k+1} \delta^{ab} \Phi_{,bc} dE^a dE^c \,.
  \label{gdown}
\ee
The function $\Phi=\Phi(E^a)$ represents the fundamental equation of classical thermodynamics \cite{callen}. 
Each thermodynamic system is given by a particular fundamental equation which, in turn, determines a particular 
metric $g$ for ${\cal E}$. In this manner, GTD provides a particular equilibrium space for each thermodynamic 
system. In classical thermodynamics, fundamental equations are usually obtained in the laboratory by applying 
empirical methods \cite{callen}. In GTD, we can use a theoretical approach to obtain new fundamental equations 
\cite{vqs10}. If we suppose that the equilibrium space ${\cal E}$ is an extremal subspace of ${\cal T}$, 
i.e., the action $I=\int \sqrt{|{\rm det}( g) |}\, d^n E$ satisfies the variational principle $\delta I =0$, we 
obtain the Nambu-Goto differential equations 
\be
  \frac{1}{\sqrt{|{\rm det}( g) |}}\left(\sqrt{|{\rm det}( g) |}\,\, g^{ab}Z^A_{,a}\right)_{,b} + 
  \Gamma^A_{\ BC} Z^B_{,b}Z^C_{,c} g^{bc} =0 \,,
  \label{meqg}
\ee
where $\Gamma^A_{\ BC}$ are the Christoffel symbols associated with the metric $G$, which is a system of differential 
equations for $Z^A(E^a)$ and in particular for the fundamental equation $\Phi(E^a)$. Consider the special case 
$\Phi = S$ and $E^a=(U,V)$, where $S$ is the entropy, $U$ the internal energy and $V$ the volume. Then, the resulting 
differential equations contain the free function $\Lambda$ and the constant $k$ from the metric \eqref{gdown}. It is 
then possible to show \cite{vqs10} that for particular choices of $\Lambda$ and $k$ the fundamental equation (up to an 
arbitrary additive constant) reads 
\be
  S = c_1 \ln \left( U + \frac{\alpha}{V} \right) + c_2 \ln \left( V - \beta  \right) \,,
  \label{gvdw}
\ee
where $c_1$, $c_2$, $\alpha$ and $\beta $ are real constants. If we choose $c_1 = 2/3$, $c_2=1$, $\alpha>0$ 
and $\beta >0$, we obtain the fundamental equation for the van der Waals gas. In general, however, the set of constants 
entering the fundamental equation \eqref{gvdw} are not fixed by the Nambu-Goto equations. The main object of the present 
work is to investigate the physical consequences of the above fundamental equation in the context of relativistic cosmology.

We note that from the point of view of GTD, in general the equilibrium space of the fundamental equation \eqref{gvdw} 
is curved. The presence of curvature in the equilibrium space is interpreted as representing the interaction between the 
constituents of the thermodynamic system. In the limiting case $\alpha=\beta=0$, the thermodynamic curvature vanishes, 
indicating the absence of thermodynamic interaction. For particular values of $c_1$ and $c_2$  this limiting fundamental 
equation represents an ideal gas which, from the point of view of statistical physics, is a system without thermodynamic 
interaction. However, the arbitrary constants $c_1$ and $c_2$ do not affect this property. As we will see, this 
simple generalization of the ideal gas plays an important role in relativistic cosmology.

In this work, we consider the consequences of the presence of thermodynamic interaction in cosmology. We will 
show that with nonzero $\alpha$ and $\beta$, and a particular choice of $c_2/c_1$, it is possible to describe an 
inflationary period in the early Universe.

This work is organized as follows. In \secref{sec:notd}, we analyse the simple case of a fundamental equation without 
interactions, and show that such systems can be used to describe the standard cosmological evolution. In \secref{sec:yestd}, 
we consider an interacting fluid and determine the parameters of the model necessary to achieve an inflationary 
period. In \secref{eq:thbehav}, we briefly comment on some of the thermodynamic properties of the fluid, before we conclude 
our work in \secref{eq:sum}.

Throughout the paper we use geometric units with $G=c=\hbar=k_{_B}=1$.

%%%%%%%%%%%%%%%%%%%%%%%%%%%%%%%%%%%%%%%%%%%%%%%%%%%%%%%%%%%%
%%%%%%%%%%%%%%%%%%%%%%%%%%%%%%%%%%%%%%%%%%%%%%%%%%%%%%%%%%%%%
\section{Cosmology without thermodynamic interaction}
\label{sec:notd}

Let us consider the Friedmann-Lema\^itre-Robertson-Walker (FLRW) metric, 
\be
  ds^2 = -dt^2 + a^2(t)\left(dr^2 + r^2 d\theta^2 + r^2 \sin^2\theta d\phi^2\right) \,,
\ee
which, in connection with a perfect fluid energy-momentum tensor, yields the Friedmann equations
\be
  \frac{\dot a^2}{a^2} = \frac{8\pi}{3} \rho \ , \quad \frac{\ddot a}{a} = -\frac{4\pi}{3}(\rho + 3 P) \,,
  \label{freq}
\ee
where we assume zero spatial curvature in accordance with observations. To integrate the above differential 
equations we need an additional equation, which we take to be the non-interacting limit of the fundamental 
equation \eqref{gvdw}, i.e., 
\be
  S= c_1 \ln U + c_2 \ln V \,,
  \label{gig}
\ee
following from GTD. For any fundamental equation to be physically relevant, it must satisfy the first law 
of thermodynamics, which in this case reads $T dS= dU + P dV$. This leads to the equations of state, 
\begin{equation}
  \frac{1}{T} = \frac{\partial S}{\partial U} \quad \mathrm{~and~} \quad \frac{P}{T} = \frac{\partial S}{\partial V} \,.
\end{equation}
A straightforward calculation shows that 
\be
  P = \frac{c_2}{c_1} \frac{U}{V} = \frac{c_2}{c_1} \rho \,,
\ee
i.e., a barotropic equation of state. Different choices of the ratio $c_2/c_1$ correspond 
to different eras of the evolution of the Universe. For instance, $c_2/c_1=1/3$ describes a radiation-dominated 
Universe, $c_2=0$ describes a dust-dominated Universe, and $c_2/c_1= -1$ corresponds to dark energy. We see that the 
fundamental equation contains all the epochs of the standard cosmological model as particular cases for specific 
choices of $c_2/c_1$. In all cases, \eqeqref{gig} represents the fundamental equation of the standard cosmological model. 
This simple fact allows us to interpret the Universe as a thermodynamic system whose properties are entirely described 
by a single fundamental equation. 

Other generalizations of this fundamental equation have been used to describe dark fluids \cite{abcq12} representing 
both matter and dark energy simultaneously. In a different context \cite{gib}, this way of deriving cosmological models 
has been found useful as well and the systems described by GTD fundamental equations were called GTD fluids.

%%%%%%%%%%%%%%%%%%%%%%%%%%%%%%%%%%%%%%%%%%%%%%%%%%%%%%%%%%%%%%%%%
%%%%%%%%%%%%%%%%%%%%%%%%%%%%%%%%%%%%%%%%%%%%%%%%%%%%%%%%%%%%%%%%%
\section{Cosmology with thermodynamic interaction}
\label{sec:yestd}

From the fundamental equation for the interacting fluid \eqref{gvdw}, the corresponding equations of state 
can be calculated, giving the temperature of the fluid as 
\begin{equation} \label{eq:temp}
  T = \frac{U}{c_1} + \frac{\alpha}{c_1 V} \,,
\end{equation}
whereas the pressure reads 
\begin{equation} \label{eq:pressureorig}
  P = \frac{c_2 U V^2 + \alpha \left[ \beta  c_1 + \left( c_2 - c_1 \right) V \right] }{c_1 V^2 \left( V - \beta  \right)} \,.
\end{equation}
We now introduce an energy density as $\rho = U/V$ and parametrize the volume as a function of the scale factor as 
$  V  = V_0 \,a^3$ so that the pressure becomes a function of $\rho$ and $a$. This parametrization uses the standard 
convention of cosmology that the scale factor at the current time $t_0$ is $a(t_0) = 1$, and thus $V_0$ is the volume 
of the Universe at current time. Furthermore, the continuity equation, 
\begin{equation}
  \dot{\rho} + 3 \frac{\dot{a}}{a} \left( \rho + P\right) = 0 \,,
\end{equation}
becomes a differential equation for $\rho$ and $a$ which can be solved explicitly to give 
\begin{equation} 
  \rho = K \frac{ \left(a^3 V_0 - \beta \right)^{-\frac{c_2}{c_1}} }{a^3}-\frac{\alpha}{a^6 V_0^2} \,,
\label{eq:rho}
\end{equation}
where $K$ is an integration constant. It is always possible to choose the values of the constants $K$ and $\alpha$
such that the energy density is positive. Moreover, by fixing $c_2/c_1$, we can obtain in principle any polynomial 
dependence of the density of the kind 
\be
  \rho_{\mathrm{inf}} \sim \frac{1}{a^m} \,.
\ee
Combinations of terms with different powers are also possible. By choosing $c_2/c_1$ 
appropriately, we can thus obtain a large number of models with inflationary behavior, for instance, a cosmological 
constant--type term. However, it is also possible to achieve a period of strong expansion with an appropriate number 
of e-folds that is \emph{not} an exact inflationary term, i.e. not a constant. This can be done with the choice 
$c_2/c_1 = -8/9$, under the assumption that the constant $\beta$ is small. Indeed, using this value in \eqref{eq:rho} 
and expanding the first term for small values of $\beta$, we obtain the density 
\begin{equation} \label{eq:rhofirstapprox}
  \rho \simeq \frac{ K  V_0^{8/9}}{a^{1/3}} - \frac{8\beta  K}{9 V_0^{1/9} a^{10/3}} - \frac{\alpha}{a^6 V_0^2} \,.
\end{equation}
The first term has the exponent $m=1/3$, and can produce the appropriate amount of e-foldings, if it is the 
dominating contribution to the density during the inflationary regime. Indeed, neglecting the last two terms in 
\eqref{eq:rhofirstapprox} for the duration of inflation, we can calculate the number of e-foldings from the inflationary 
density 
\be
  \rho (a) \simeq \frac{ K  V_0^{8/9}}{a^{1/3}} \equiv \rho_{\mathrm{inf}} (a) 
\ee
as follows. The integration of the first Friedmann equation yields the scale factor and the Hubble parameter 
\begin{equation} \label{eq:scale}
  a = \left( \frac{2\pi}{27} K V_0^{8/9} \right)^{1/6} t^6 \,,  \ \   H = \frac{6}{t} \,.
\end{equation}
The number of e-foldings then can be calculated as 
\begin{equation} \label{eq:N}
  N = \int H dt = 6 \ln \left( \frac{t_f}{t_i} \right) \,,
\end{equation}
where $t_i$ and $t_f$ are the times of the beginning and end of inflation, usually estimated to be $t_i = 10^{-36} 
\mathrm{s}$ and $t_f = 10^{-32} \mathrm{s}$; in this case, we obtain $  N = 6 (-32 + 36) \ln 10 \simeq 55$, which is 
an appropriate amount of e-foldings.

The assumption that the inflationary density is dominated by the first term given in \eqref{eq:rhofirstapprox} puts 
constraints on the possible values of the constants $\alpha$ and $\beta $, for instance, by requiring that the absolute 
value of each of the two additional terms in \eqref{eq:rhofirstapprox} is much smaller than the absolute value of the 
inflationary term, i.e. 
\begin{eqnarray} 
  && \left| \alpha \right| \ll K V_0^{26/9} a_i^{17/3} =: \alpha_c \,, \label{eq:condInf1}\\
  && |\beta | \ll \frac{9}{8}\, V_0 \, a_i^3 =: \beta _c \,, \label{eq:condInf2}
\end{eqnarray}
where $a_i = a(t_i)$. 
Here the signs of $\alpha$ and $\beta $ do not matter, since only their absolute values have to be  negligible. From 
the fundamental equation of GTD \eqref{gvdw}, it follows that the constant $\beta$ should be related with some 
characteristic volume. Therefore we assume that $\beta $ be positive. On the other hand, $\alpha$ usually is the 
interaction constant between the gas constituents, and could in principle 
be both positive and negative. However, if we choose $\alpha$ to be negative, and let the two terms proportional to 
$\alpha$ and $\beta$ cancel each other exactly at the beginning of inflation, we end up with the condition 
\begin{equation} \label{eq:condInfStrong}
 \left| \alpha \right| = \frac{8}{9} \beta K V_0^{17/9} a_i^{8/3} \ll \alpha_c \,,
\end{equation}
which, with the condition on $\beta $, leads to the same condition on $\alpha$ as in the previous case. This would 
mean that inflation starts off very cleanly, since the density at the beginning of inflation is exactly composed by 
the inflationary term only, and the contribution of the other two densities will first slightly grow, before the terms 
dilute away faster than the inflationary term in the course of inflation due to their dependence on the scale factor. 
All dynamics and physical properties will thus be determined in terms of a small parameter 
\begin{equation} \label{eq:epsilon}
  \epsilon = \frac{\beta }{\beta _c} = -\frac{|\alpha|}{\alpha_c} \,.
\end{equation}
Moreover, the present inflationary model contains two additional parameters, namely, $V_0$ and $K$.

To get a quantitative grip on the involved constants, we consult what is considered the standard picture of cosmology. 
Inflation in our model lasts for about $55$ e-foldings, i.e. during inflation the Universe expands for a factor of roughly 
$e^{55} \simeq 7 \cdot 10^{23}$ times. After that, the Universe grows for further $10^{30}$ times during the periods of 
radiation and matter dominance \cite{2003Line}. We further know that the current observable Universe has a diameter of about 
$l_0 \simeq 10^{26} \,\mathrm{m}$, i.e. a volume of about $V_0 \simeq 10^{78} \,\mathrm{m^3}$. 
The diameter of the Universe at the beginning of inflation was thus $ l_i = l_0 / (10^{30} e^{55}) \simeq 10^{-28} \,\mathrm{m}$. 
Using $V=V_0 a^3$ and the convention $a(t_0) = 1$, we can thus determine the scale factor at the beginning of inflation as 
$a_i = l_i/l_0 \simeq 10^{-54}$. Combining these numbers, we obtain
\be
  \beta _c = \frac{9}{8} V_0 a_i^3 \simeq 10^{-84} \, \mathrm{m^3} \,,
\ee
i.e. $\beta_c$ equals about the volume of the Universe at the onset of inflation -- a small number, but still much larger 
than the Planck volume, $l_p^3 \simeq 10^{-105} \, \mathrm{m}^3$. For $\alpha_c$, we have to estimate the value of $K$, 
which can be fixed requiring that the energy scale at the onset of inflation was of the order of the GUT scale of about 
$10^{16}\, \mathrm{GeV}$, 
\begin{equation}
  \rho_{\mathrm{inf}} \left(t_i\right) = \frac{K V_0^{8/9}}{a_i^{1/3}} = \frac{10^{16} \, \mathrm{GeV}}{l_i^3} \simeq 
  \frac{10^6 \mathrm{J}}{10^{-84} \, \mathrm{m^3}} = 10^{89} \, \mathrm{\frac{J}{m^3}} \,.
\end{equation}
The constant $K$ then is 
\begin{equation}
   K = \frac{\rho_{\mathrm{inf}} \left(a_i\right) a_i^{1/3}}{V_0^{8/9}} \simeq 2\cdot 10^{2} \, \mathrm{J m^{-17/3}} \,.
\end{equation}
The peculiar unit of $K$ is owed to the requirement that the inflationary density has the unit of energy density, and 
leads to 
\begin{equation}
  \alpha_c \simeq 10^{-78} \, \mathrm{J m^3} \,.
\end{equation}

Ultimately, it is instructive to see how fine-tuned the value of $c_2/c_1$ has to be in order to achieve an appropriate 
amount of e-foldings. We use the expression for the density \eqref{eq:rho}, without specifying $c_2/c_1$, and expand it 
for small $\beta$. Neglecting the $\alpha$-- and $\beta$--terms, the inflationary density in the general case results in 
\begin{equation}
  \rho_{\mathrm{inf}} \simeq \frac{ k }{V_0^{c_2/c_1}} \frac{1}{a^{m}} \,,
    \quad  m = 3 + 3 \frac{c_2}{c_1} \,.
\end{equation}
Using \eqref{eq:N} for the number of e-foldings, we can calculate the values of $m$ and $N$ for a range of choices 
of $c_2/c_1$: 
\begin{eqnarray}
  \frac{c_2}{c_1}\Bigg|_1 = -0.912 \quad \Longrightarrow \quad m_1 \simeq 0.263\,, \quad N_1 \simeq 70 \,, \nonumber\\
  \frac{c_2}{c_1}\Bigg|_2 = -0.898 \quad \Longrightarrow \quad m_2 \simeq 0.307\,, \quad N_2 \simeq 60 \,, \\
  \frac{c_2}{c_1}\Bigg|_3 = -0.877 \quad \Longrightarrow \quad m_3 \simeq 0.368\,, \quad N_3 \simeq 50 \,. \nonumber 
\end{eqnarray}
The value of $m=1/3$ leading to $N \simeq 55$ corresponding to $c_2/c_1 = -8/9$ lies somewhere between $N_2$ and $N_3$.

%%%%%%%%%%%%%%%%%%%%%%%%%%%%%%%%%%%%%%%%%%%%%%%%%%%%%%%%%%%%%%%%%%%%%%%%%%%%%%%%%%%%%%%%%%%%%%%%%%%%%%%%%%%%%%%%%%
\section{Thermodynamic behaviour}
\label{eq:thbehav}

Using the expressions for the volume $V=V_0a^3$ and the density \eqref{eq:rho} for  $c_2/c_1 = -8/9$,
and carrying out an expansion for small $\beta$, the temperature \eqref{eq:temp} becomes
\begin{equation}
   T(a) \simeq  \frac{K}{c_1}  a^{8/3} \left( V_0^{17/9} - \frac{8 \beta  V_0^{8/9}}{9} \frac{1}{a^3} \right) \,.
\end{equation}
Using the definition \eqref{eq:epsilon} for $\epsilon$, and rewriting the scale factor as multiples of its value at 
the onset of inflation, 
\begin{equation} \label{eq:x}
  a(t) = a_i x \,,
\end{equation}
the temperature can be reexpressed as 
\begin{equation} \label{eq:Tinf}
   T(x) \simeq  \frac{K}{c_1} V_0^{17/9} a_i^{8/3} x^{8/3} \left( 1 - \epsilon x^{-3} \right) \,.
\end{equation}
Without loss of generality we can assume that $c_1$ is positive. Then, the temperature is positive as long as the 
expression in the bracket is positive, i.e. as long as $x$ keeps growing and $\epsilon$ is small, which is one of the 
conditions which already ensures the dominance of the inflationary term, \eqref{eq:condInf2}. Consequently, the same 
condition that must be satisfied in order to have clean inflation also guarantees that the temperature is positive at 
all times during inflation.  

From the general expression for the  pressure \eqref{eq:pressureorig} and the density \eqref{eq:rho} for the special 
inflationary case  $c_2/c_1 = -8/9$ and small values of $\beta$, 
%we obtain 
%\begin{equation} 
%  P(a) \simeq -\frac{\alpha}{a^6 V_0^2} - \frac{8 K}{9} V_0^{8/9} a^{-1/3} 
 %     - \frac{8 \beta K}{81} V_0^{-1/9} a^{-10/3} \,.
%\end{equation}
and using \eqref{eq:epsilon} and \eqref{eq:x}, we get 
\begin{equation} \label{eq:PvdW}
  P_{inf}(x) = \frac{\rho_{\mathrm{inf}}(a_i)}{x^{1/3}} \left[ - \frac{8}{9} - \frac{\epsilon}{9} x^{-3} 
    + \epsilon x^{-17/3} \right] \,.
\end{equation}
We see that for  small values of  $\epsilon$, the pressure is negative and nearly that of a cosmological constant
with barotropic factor $ \omega_{inf} = - 8 /9$. During inflation, $x$ will increase, which implies that the terms 
proportional to epsilon will dilute away much faster than the leading dependence on $x$. The pressure is thus always 
negative during inflation.

%%%%%%%%%%%%%%%%%%%%%%%%%%%%%%%%%%%%%%%%%%%%%%%%%%%%%%%%%%%%%%%%%%%%%%%%%%%%%%
%%%%%%%%%%%%%%%%%%%%%%%%%%%%%%%%%%%%%%%%%%%%%%%%%%%%%%%%%%%%%%%%%%%%%%%%%%%%%%%
\section{Summary}
\label{eq:sum}
In this work, we have investigated the physical properties of a particular thermodynamic fundamental equation that is 
obtained from the GTD formalism, in the framework of relativistic cosmology. It represents the entropy of a thermodynamic 
system that explicitly depends on the internal energy and volume. In addition, it contains four real parameters 
$c_1$, $c_2$, $\alpha$ and $\beta$ which  enter the corresponding equation of state. For a particular choice of these 
parameters, the fundamental equation corresponds to that of a van der Waals gas. 

Furthermore, if we assume that this equation of state can be applied to the entire Universe, we construct a cosmological 
model that describes the  evolution of the Universe. Indeed, based on the interpretation of thermodynamic interaction 
as the curvature of the equilibrium space, we consider two  different cosmological models, with and without interaction. 
The cosmological model without thermodynamic interaction, which corresponds to the limiting case $\alpha=\beta=0$,  turns 
out to be equivalent to the standard $\Lambda$CDM model. On the other hand, the resulting cosmology in the presence of 
thermodynamic interaction has been shown to reproduce the main features of inflation, namely, the number of e-foldings 
($N\approx 55)$, and is consistent with commonly assumed parameters as the initial time ($t_i\approx 10^{-36} \,\mathrm{s}$) 
and the final time ($t_f\approx 10^{-32} \,\mathrm{s}$). This inflationary model is valid only for the particular 
value $c_2/c_1=-8/9$ and under the condition that the parameters  $\alpha$ and $\beta$ are small. Evaluating these parameters 
shows that $\beta$ corresponds to the volume of the Universe at the beginning of inflation and turns out to be 
$\approx 10^{-84}\, \mathrm{m^3}$. On the other hand, the ratio $\alpha/\beta$ determines the internal energy of the Universe 
at the beginning of inflation. The interaction constant $\alpha$ turns out to be small and $\approx 10^{-78} \,\mathrm{J m^3}$. 
These properties can be considered predictions of our cosmological model. 

By relating $\alpha$ and $\beta$, we have developed a model for inflation depending on only one semi-free 
parameter, i.e. $\epsilon$ (semi-free because it has to obey certain limits), after fixing the remaining constants $V_0$ and 
$K$ to physically reasonable values expected from an inflationary model, and choosing the value of the parameter $c_2/c_1$ 
determining the equation of state of the fluid. 
The inflationary model derived in this work is thus unique in the sense that all the parameters entering the 
fundamental equation must be fixed or limited by critical values in order to reproduce the main features of inflation. 
A fine-tuning, however, is not 
necessary, as we have seen. This implies that the GTD inflationary model presented in this work is not an element of a 
family of models in which many possible scenarios are consistent with observations, but instead predicts a specific scenario 
which can be compared with observations to determine its validity. 

One way to test the model are cosmological perturbations, a very important aspect of any inflationary model. A detailed 
analysis of this issue is, 
however, beyond the scope of the present work. Preliminary analysis have shown that a nearly scale-invariant power 
spectrum of primordial perturbations can be achieved within the framework of our proposal. However, since the model 
is constructed not from the dynamics of a scalar field but from the thermodynamic behavior of an interacting system, the 
conventional calculation of a primordial power spectrum from the quantum fluctuations of the scalar field does not 
apply here. Instead, one has to consider perturbations in the fluid limit of the Einstein equations, or some alternative 
thermodynamic description of perturbations in the fluid. A detailed analysis of the perturbation spectrum will be 
presented in a more extensive complementary publication, along with a thorough examination of the properties of the 
inflationary fluid, considering thermodynamic response functions, possible phase transitions and critical points.

\section*{Acknowledgements}
CG was supported by an UNAM postdoctoral fellowship program. This work has been supported by the UNAM-DGAPA-PAPIIT, Grant 
No. IN111617.

\end{document}